\documentclass[prd,aps, preprintnumbers, showpacs,twocolumn, nofootinbib,superscriptaddress,notitlepage]{revtex4-1}
\usepackage[normalem]{ulem}
\usepackage{epsfig}
\usepackage{amsfonts}
\usepackage{amsmath}
\usepackage{slashed}
\usepackage{graphicx}
\usepackage{color}
\usepackage{mathtools}
\usepackage{subfigure}
\usepackage{stmaryrd}
\usepackage{amssymb,psfrag}
\usepackage{pifont}
\allowdisplaybreaks[4]

\newcommand{\beq}{\begin{eqnarray}}
\newcommand{\eeq}{\end{eqnarray}}

\newcommand{\nn}{\nonumber}

\def\slash#1{#1 \hskip-0.45em /}

\DeclareRobustCommand{\app}[1]{Appendix~\ref{sec:#1}}

\allowdisplaybreaks

\begin{document}

\title{Accessing the HQET B-Meson Shape Function from a LaMET Quasi-Shape Function}

\author{Ao-Sheng Xiong}
\email{xiongash21@lzu.edu.cn}
\affiliation{Frontiers Science Center for Rare Isotopes, and School of Nuclear Science and Technology, Lanzhou University, Lanzhou 730000, China}

\author{Jun Ding}
\affiliation{Frontiers Science Center for Rare Isotopes, and School of Nuclear Science and Technology, Lanzhou University, Lanzhou 730000, China}

\author{Ji Xu}
\email{xuji@lzu.edu.cn, corresponding author}
\affiliation{Frontiers Science Center for Rare Isotopes, and School of Nuclear Science and Technology, Lanzhou University, Lanzhou 730000, China}

\author{Jun Zeng}
\email{zengj@hainnu.edu.cn, corresponding author}
\affiliation{College of Physics and Electronic Engineering, Hainan Normal University, Haikou 571158, Hainan, China}

\author{Shuai Zhao}
\email{zhaos@tju.edu.cn, corresponding author}
\affiliation{Department of Physics and Center for Joint Quantum Studies, School of Science, Tianjin University, Tianjin 300350, China}

\begin{abstract}
The shape function and the light-cone distribution amplitude of heavy meson jointly characterize the nonperturbative structure of the heavy meson on the light-cone, with the former being essential for theoretical predictions of inclusive decays and the latter for exclusive decays. While first-principles lattice QCD results for the heavy meson LCDA have become available in recent years, lattice results for the shape function remain absent. In this work, we establish a two-step factorization scheme---known as the HQLaMET framework---for computing the $B$-meson shape function on the lattice, which fully disentangles the effects of the disparate scales $P_B^z$, $m_b$, and $\Lambda_{\textrm{QCD}}$. For illustration, starting from a phenomenological model for the shape function in HQET, we provide a graphical presentation of the entire procedure of this framework. The results of the current work lay the foundation for nonperturbative lattice QCD determinations of the shape function in the near future.
\end{abstract}

\maketitle

\section{Introduction}
\label{sec:introduction}

The shape function (SF) and the light-cone distribution amplitude (LCDA) of a heavy-light meson defined in heavy quark effective theory (HQET) are fundamental nonperturbative inputs underlying precision studies of inclusive and exclusive $B$-meson decay processes, respectively \cite{Neubert:1993um,Grozin:1996pq}. Inclusive decays, such as the semileptonic process $B\to X_u \ell\bar{\nu}$ and the radiative process $B\to X_s \gamma$, are of great importance for the determination of the CKM matrix element $|V_{ub}|$ and for probing new physics beyond the Standard Model at $B$ factories. Notably, the SF governs the photon energy spectrum and the lepton invariant-mass spectrum in these channels \cite{Neubert:1993ch,Bosch:2004th,Zhang:2024rmb,Neubert:2004sp}. In the exclusive decays, the LCDAs are indispensable ingredients in the factorization theorems~\cite{Beneke:1999br,Beneke:2000wa,Beneke:2001ev,Becher:2005bg,Na:2015kha,Gao:2021sav,Shi:2026mjb,Deng:2024dkd}. It is well known that the discrepancy between this CKM matrix element extracted from inclusive and exclusive $B$ decays---known as the ``$|V_{ub}|$ puzzle''---has persisted for many years \cite{ParticleDataGroup:2024cfk,HFLAV:2022esi,Cao:2023rku}. A simultaneous improvement in our understanding of both the SF and the LCDA represents a critical step toward resolving this discrepancy. In addition, any theoretical insight into the SF that could sharpen the analysis of inclusive $B$ decays would also enhance analysis sensitivity to potential signatures of physics beyond the Standard Model \cite{Ligeti:2008ac,Shi:2023riy,Misiak:2006zs}.

However, both the SF and the LCDA are intrinsically nonperturbative objects and have resisted first-principles determination for more than three decades since their introduction. The difficulty is fundamental: these quantities are defined through matrix elements of non-local operators along the light-cone, which obstructs the standard lattice QCD approaches. As a result, phenomenological analyses have long relied on model parametrizations with poorly constrained SF and LCDA \cite{Braun:2003wx,Lee:2004ja,Kawamura:2001jm,Li:2006jb,Braun:2019wyx,Gao:2019lta,Gao:2024vql,Chai:2022ptk,Balzereit:1998yf,Bauer:2003pi,Aglietti:1999ur,Bauer:2000ew,Hua:2020usv}, and the selection of specific models inevitably introduces uncertainties and biases. The model dependence of these two quantities constitutes one of the dominant limitations in precision predictions for $B$-meson decays. Therefore, a significant task in heavy flavor physics is to establish methods for accessing these quantities in model-independent manner.

The development of large-momentum effective theory (LaMET) and related approaches has opened a new path for computing light-cone quantities from lattice QCD \cite{Ji:2013dva,Ji:2014gla,Ji:2020ect,Zhao:2022qym,Radyushkin:2022qvt,Zhao:2020bsx,Zhang:2018diq}. LaMET establishes a framework in which a quasi-distribution, defined through equal-time spatial correlators, can be related to the corresponding light-cone distribution through a factorization formula. Extensive expertise has been accumulated in applying the LaMET formalism to compute parton distribution functions of nucleons and distribution amplitudes of light hadrons \cite{Cichy:2018mum,Xu:2018mpf,Liu:2018tox,Liu:2019urm,Hua:2020gnw,Xu:2022krn}. However, its application to heavy mesons remains limited, primarily because preserving the heavy quark mass in lattice simulations notably increases the complexity of computation \cite{Wang:2019msf,Hu:2023bba,Hu:2024ebp,Wang:2025uap,Ishaq:2019dst,Beneke:2023nmj}.

A decisive advance was made in the past few years when the two-step factorization scheme---the heavy quark large momentum effective theory (HQLaMET) framework---was introduced for simulating the $B$-meson LCDA on the lattice \cite{Han:2024fkr,LatticeParton:2024zko}. In this framework, the first step connects the lattice computable quasi-distribution amplitude to the QCD LCDA through LaMET matching, while the second step converts the QCD LCDA into the HQET LCDA via a factorization formula that separates the heavy quark mass scale from the soft scale $\Lambda_{\textrm{QCD}}$. The preliminary findings within this scheme are qualitatively consistent with phenomenological models and with experimental constraints from $B\to\gamma\ell\nu_\ell$ \cite{Belle:2018jqd}, demonstrating a promising prospect. Very recently, this program has been elevated from proof-of-concept to precision determination through a precise lattice QCD calculation that incorporates systematic error control \cite{LPC:2026vyv,LPC:2026ffe}. The HQLaMET framework has enabled rapid progress in the first-principles lattice calculation of LCDAs of heavy mesons over the past two years.

Despite the fact that both SF and LCDA were introduced in the same period (the 1990s), no first-principles lattice calculation of the $B$-meson SF has been performed to date. In a similar spirit, this work aims to establish a complete two-step factorization scheme for the $B$-meson SF, thereby advancing lattice simulations in the near future. We demonstrate that the HQLaMET framework, originally formulated for LCDAs, is equally applicable to SF computations. Specifically, we derive the one-loop matching coefficient between the quasi-SF in LaMET and the QCD SF, which constitutes the first step of this framework. Combined with the second step matching between the QCD SF and HQET SF which has been given in our earlier work \cite{Han:2024fkr}, the complete HQLaMET framework is realized. We also derive a differential equation governing the dependence of the QCD SF on its heavy quark mass, and solve this equation analytically in momentum space. This is a result of particular interest for lattice calculations where simulations at the physical $b$ quark mass may not be feasible \cite{Wang:2024wwa}. Furthermore, starting from a phenomenological model for the HQET SF, we graphically illustrate the complete workflow of this framework, which will serve as a reference for future lattice simulations. The results presented in this paper provide a realistic method for enabling reliable first-principles predictions of the SF, with a systematic way to access its nonperturbative form under full theoretical control.

This paper is organized as follows. In Sec.\,\ref{sec:definitions}, we present the definitions of the $B$-meson SFs in LaMET, QCD, and HQET, followed by the factorization formulas that relate them and the tree-level matching results. In Sec.\,\ref{sec:firststep}, we calculate the one-loop matching coefficient between the quasi-SF and the QCD SF. In Sec.\,\ref{sec:secondstep}, we present the factorization between the QCD and HQET SFs at next-to-leading order. In Sec.\,\ref{sec:massevolution}, we derive the mass dependence equation and its solution. In Sec.\,\ref{sec:numerical}, we provide a numerical illustration of the entire HQLaMET framework. We conclude with a summary and outlook in Sec.\,\ref{sec:summary}. The Appendix collects supplementary results of one-loop calculations.

\section{Definitions and tree-level matching}
\label{sec:definitions}

In this section, we present the definitions of the $B$-meson SFs in LaMET, QCD, and HQET, followed by the factorization formulas that relate them. We then display the matching coefficients at tree level, which serve as preliminaries for the one-loop calculation in the next section.

\subsection{$B$-meson shape functions defined in LaMET, QCD, and HQET}
\label{sec:def_SFs}

We first introduce the quasi-SF, which is both calculable on the lattice and amenable to matching onto the QCD SF. In the HQLaMET framework, the $B$-meson quasi-SF is defined as \cite{Wang:2025uap}
\begin{eqnarray}\label{eq:def_quasi_SF}
  && \tilde{S}^{\textrm{LaM.}}(x,\mu) = \int \frac{d z \,P_B^z}{2\pi} e^{-i x z P_B^z} \nn\\
  &&\quad \times \frac{\langle B(P_B)|(\bar{b}W)(z) \frac{\gamma^t}{2} (W^\dagger b)(0)|B(P_B)\rangle }{\langle B(P_B)|(\bar{b}W)(0) \frac{\gamma^t}{2} (W^\dagger b)(0)|B(P_B)\rangle } \,,
\end{eqnarray}
where the Dirac matrix $\gamma^t \equiv \gamma\cdot n_t$ with the spatial coordinates:
\begin{eqnarray}
  n_{t\mu}=(1,0,0,0) \,, \quad n_{z\mu}=(0,0,0,1) \,. \nn
\end{eqnarray}
 $P_B$ denotes the momentum of $B$-meson, with $P_B^z$ being its spatial $z$-component; the Wilson line $W$ is along the $z$-direction, ensuring gauge invariance. The quasi-SF is constructed from the spatial correlation function of two heavy quark fields defined in QCD, and we work in a Lorentz-boosted frame of the $B$-meson in which $P_B^z \gg m_Q$. Unlike the QCD SF defined below, which is invariant under a boost along the $z$ direction, the quasi-SF changes dynamically under such a boost, encoded in its nontrivial dependence on $P_B^z$.

Another quantity that captures the nonperturbative structure of the heavy meson is the SF defined in QCD, which includes short-distance physics at energy scales near the heavy quark mass~\cite{Wang:2025uap}. The QCD SF is defined as
\begin{eqnarray}\label{eq:def_QCD_SF}
  && S^{\textrm{QCD}}(y,\mu) = \int \frac{dz^-\, P_B^+}{2\pi} e^{-iyz^- P_B^+} \nn\\
  &&\quad \times \frac{\langle B(P_B)|\bar{b}(0) \,\slash{n}_+ \, W(0,z) \, b(z)|B(P_B)\rangle }{\langle B(P_B)|\bar{b}(0) \, \slash{n}_+ \, b(0)|B(P_B)\rangle } \,.
\end{eqnarray}
Here $z^2=0$ denotes the lightcone separation and $0 \leq y \leq 1$ signifies the lightcone momentum fraction of the $b$ quark inside the hadron. The Dirac matrix $\Gamma=\slash{n}_+$ is chosen such that the above definition is analogous to that of the quark PDFs of a nucleon \cite{Izubuchi:2018srq}. The notations for lightcone coordinates are
\begin{eqnarray}
  n_{+\mu}=\frac{1}{\sqrt{2}}(1,0,0,1) \,, \quad n_{-\mu}=\frac{1}{\sqrt{2}}(1,0,0,-1) \,. \nn
\end{eqnarray}

Lastly, we introduce the SF defined in HQET, which is the ultimate nonperturbative target of this framework. The HQET SF is defined in terms of the forward matrix elements of non-local operators along the lightcone, i.e., a hadron-to-hadron matrix element in the infinite quark mass limit~\cite{Neubert:1993um}:
\begin{eqnarray}\label{eq:def_HQET_SF}
  && S^{\textrm{HQET}}(\omega,\mu) =  \int_{-\infty}^{+\infty} \frac{dt\,v^+}{2\pi} e^{i \omega  t v^+} \nn\\
  &&\quad \times \frac{\langle B(v) | \bar h_v(0) \, W(0,tn_+) \, h_v(tn_+) | B(v) \rangle}{\langle B(v) | \bar h_v(0) \, h_v(0) | B(v) \rangle} \,,
\end{eqnarray}
where $h_v$ denotes the effective heavy quark field with four-velocity $v$ satisfying $v^2=1$; the Wilson line $W$ is $W(0,tn_+)=\textrm{P}\,\left\{\exp \left[ -i g_s \int_0^{tn^+} ds \, n_+\!\cdot\!A(sn_+) \right]\right\}$ that ensures gauge invariance; $|B(v)\rangle$ is the $B$-meson state in HQET. We denote $v^{+}\equiv n_{+} \!\cdot\! v$ and $v^{-} \equiv n_{-} \!\cdot\! v$ for convenience.

The SF describes the distribution of residual lightcone momentum $\omega$ of the $b$ quark inside the $B$ meson, and has support for $-\infty < \omega \leq \bar\Lambda$ with $\bar\Lambda = m_B-m_b$~\cite{Bosch:2004th}. Physically, it determines the ``Fermi motion'' of the $b$ quark inside the $B$-meson, which is analogous to the motion of nucleons within an atomic nucleus or the role of parton distribution functions in a nucleon.

\subsection{Factorization formulas}
\label{sec:fac_formulas}

The quasi-SF and the QCD SF exhibit distinct ultraviolet behaviors. Nevertheless, these two quantities share the same infrared behavior, which necessitates a factorization formula.

Starting from the quasi-SF, the first step integrates out the large spatial momentum $P_B^z$ and yields the QCD SF \cite{Ji:2013dva}:
\begin{eqnarray}\label{eq:fac1}
  \tilde{S}^{\textrm{LaM.}}(x,\mu) = \int_0^1 dy\, Z_\Gamma(x,y,\mu) \, S^{\textrm{QCD}}(y,\mu) \,.
\end{eqnarray}
This factorization separates the large momentum scale $P_B^z$ from $m_b$ and $\Lambda_{\textrm{QCD}}$. The matching coefficient $Z_\Gamma(x,y,\mu)$ can then be determined reliably through the perturbative matching procedure.

In the second step, the QCD SF is related to the HQET SF through another factorization formula that separates the heavy quark mass scale $m_b$ from the nonperturbative scale $\Lambda_{\textrm{QCD}}$. It is convenient to consider the QCD SF separately in different kinematic regions. The region where $1-y \sim \mathcal{O}(\Lambda_{\textrm{QCD}}/m_b)$ is referred to as the ``peak region'', characterized by large momentum fractions of the heavy quark. The region $1-y \sim \mathcal{O}(1)$ is termed the ``tail region'', in which the behavior of $S^{\textrm{QCD}}(y,\mu)$ can be computed perturbatively. The factorization formula takes the form~\cite{Wang:2025uap}:
\begin{eqnarray}\label{eq:fac2}
  && S^{\textrm{QCD}}(y,\mu) = \nn\\
  && \quad \begin{cases}
  \int_{-\infty}^{\bar\Lambda} d\omega \, Z_{\textrm{peak}}(y, \omega, \mu) \, S^{\textrm{HQET}}(\omega,\mu) \,, & 1\!-\!y \!\sim\!  \mathcal{O}(\lambda) \,, \\ \\
  Z_{\textrm{tail}}(y,\mu) \,, & 1\!-\!y \!\sim\! \mathcal{O}(1) \,,
  \end{cases}
\end{eqnarray}
where $\lambda = \Lambda_{\textrm{QCD}}/m_b \ll 1$. We emphasize that the tail region is fully computable in perturbation theory and is not our primary focus. The nonperturbative information resides entirely in the peak region.

The matching coefficients $Z_\Gamma(x,y,\mu)$, $Z_{\textrm{peak}}(y, \omega, \mu)$, and $Z_{\textrm{tail}}(y,\mu)$ are insensitive to infrared physics and can be determined through standard perturbative matching. Since they are free of long-distance effects, the $B$-meson state can be replaced with a free $b\bar{q}$ quark state, allowing both sides of the factorization formulas to be computed in perturbation theory. Expanding in powers of $\alpha_s$, we have:
\begin{subequations}\label{eq:expansions}
\begin{eqnarray}
  \tilde{S}^{\textrm{LaM.}}(x,\mu) &=& \tilde{S}^{\textrm{LaM.}(0)}(x,\mu) + \frac{\alpha_s C_F}{2\pi} \tilde{S}^{\textrm{LaM.}(1)}(x,\mu) \nn\\
  &&  +\mathcal{O}(\alpha_s^2) \,,\\
  S^{\textrm{QCD}}(y,\mu) &=& S^{\textrm{QCD}(0)}(y,\mu) + \frac{\alpha_s C_F}{2\pi} S^{\textrm{QCD}(1)}(y,\mu) \nn\\
  && +\mathcal{O}(\alpha_s^2) \,,\\
  S^{\textrm{HQET}}(\omega,\mu) &=& S^{\textrm{HQET}(0)}(\omega,\mu) + \frac{\alpha_s C_F}{2\pi} S^{\textrm{HQET}(1)}(\omega,\mu) \nn\\
  && +\mathcal{O}(\alpha_s^2) \,.
\end{eqnarray}
\end{subequations}
One is then able to solve the matching coefficients order by order in $\alpha_s$,
\begin{subequations}\label{eq:Z_expansions}
\begin{eqnarray}
  Z_{\Gamma}(x, y, \mu) &=& Z_\Gamma^{(0)}(x,y,\mu) +\frac{\alpha_s C_F}{2\pi}Z_\Gamma^{(1)}(x,y,\mu) \nn\\
  && +\mathcal{O}(\alpha_s^2) \,,\\
  Z_{\textrm{peak}}(y,\omega,\mu) &=& Z_{\textrm{peak}}^{(0)}(y,\omega,\mu) + \frac{\alpha_s C_F}{2\pi}Z_{\textrm{peak}}^{(1)}(y,\omega,\mu) \nn\\
  && +\mathcal{O}(\alpha_s^2) \,,\\
  Z_{\textrm{tail}}(y,\mu) &=& Z_{\textrm{tail}}^{(0)}(y,\mu) + \frac{\alpha_s C_F}{2\pi}Z_{\textrm{tail}}^{(1)}(y,\mu) \nn\\
  && +\mathcal{O}(\alpha_s^2) \,.
\end{eqnarray}
\end{subequations}
We identify the momenta of the $B$-meson and $b$ quark as $P_B^+ = m_B v^+$, $P_b^+ = m_b v^+ + k^+$, and similarly $P_B^z = m_B v^z$, $P_b^z = m_b v^z + k^z$, where $k$ denotes the residual momentum of $b$ quark with $\tilde k = k^+/v^+ = k^z/v^z$.

At tree level, the results are straightforward. For the first step, the quasi-SF and QCD SF take the forms:
\begin{subequations}
\begin{eqnarray}
  \tilde{S}^{\textrm{LaM.}(0)}(x,\mu) &=& \delta\Big( x-\frac{m_b}{m_B} - \frac{\tilde{k}}{m_B} \Big) \,, \\
  S^{\textrm{QCD}(0)}(y,\mu) &=& \delta\Big( y-\frac{m_b}{m_B} - \frac{\tilde{k}}{m_B} \Big) \,,
\end{eqnarray}
\end{subequations}
which immediately yield the tree level matching coefficient for the first step:
\begin{eqnarray}\label{eq:Z_tree}
  Z^{(0)}_\Gamma(x,y,\mu) = \delta(x-y) \,.
\end{eqnarray}
This simple result confirms that the quasi-SF reduces to the QCD SF at leading order.

For the second step, the tree level matching coefficient is \cite{Wang:2025uap}:
\begin{subequations}\label{eq:Z2_tree}
\begin{eqnarray}
  Z_{\textrm{peak}}^{(0)}(y,\omega,\mu) &=& \delta\Big( \frac{\omega}{m_B} +\frac{m_b}{m_B} -y \Big) \,, \\
  Z_{\textrm{tail}}^{(0)}(y,\mu) &=& 0 \,.
\end{eqnarray}
\end{subequations}
These tree level results are consistent with intuitive expectations: the first step pins $x$ to $y$, while the second step relates the lightcone fraction $y$ to the HQET residual momentum $\omega$ through a kinematic mapping. It is worth noting that, since the typical momentum of $\omega$ is soft, the QCD SF has support primarily in a small region near the endpoint.

\section{Matching relation between $\tilde{S}^{\textrm{LaM.}}$ and $S^{\textrm{QCD}}$ at next-to-leading order}
\label{sec:firststep}

This section presents the matching coefficient between $\tilde{S}^{\textrm{LaM.}}(x,\mu)$ and $S^{\textrm{QCD}}(y,\mu)$ at the one-loop level. Here we determine $Z_\Gamma^{(1)}(x,y,\mu)$ through the matching formula in Eq.\,(\ref{eq:fac1}). Before delving into the details, we outline several key points to help clarify the results presented below.
\begin{itemize}
  \item We regulate the ultraviolet (UV) divergences dimensionally in space-time dimension $d=4-2\epsilon$ and employ the $\overline{\textrm{MS}}$ scheme. In order to identify the relevant infrared (IR) singularities, we keep $v \cdot k$ nonzero as the IR regulator throughout the main text.

  \item It is convenient to regulate the singularity that arises when $x \to y$ by a plus distribution. We employ the plus distribution defined by
      \begin{eqnarray}\label{eq:def_plus}
        \left[F(x, y)\right]_{\oplus} &=& F(x, y) - \delta(x-y) \int_{-\infty}^{+\infty} dt \, F(t, y) \,.\nn
      \end{eqnarray}

  \item A model-independent description of the asymptotic behavior (tail region) of $S^{\textrm{HQET}}(\omega,\mu)$ for large $|\omega|$ has been given in Ref.\,\cite{Bosch:2004th}. Since the present work focuses primarily on the peak region, we omit further discussion of the tail region.

  \item As a nontrivial cross-check, we also evaluate the matching in a complementary scheme in which the heavy quark mass is set to zero and a nonzero gluon mass $m_g\neq 0$ is used as the IR regulator. The results, collected in \app{appendixA}, are found to be consistent with those obtained in the main text.
\end{itemize}

\begin{widetext}
\subsection{One-loop results for the quasi-shape function $\tilde{S}^{\textrm{LaM.}}$ in LaMET}
\label{sec:oneloop_quasi}

The one-loop Feynman diagrams consist of the heavy-quark sail graph, Wilson-line self-energy graph, box graph, and local vertex graph, as shown in Fig.\,\ref{fig:feynman}. The calculation involves the same Feynman diagram structures for all three kinds of SFs.

\begin{figure*}[ht]
  \centering
  \includegraphics[width=0.9\textwidth]{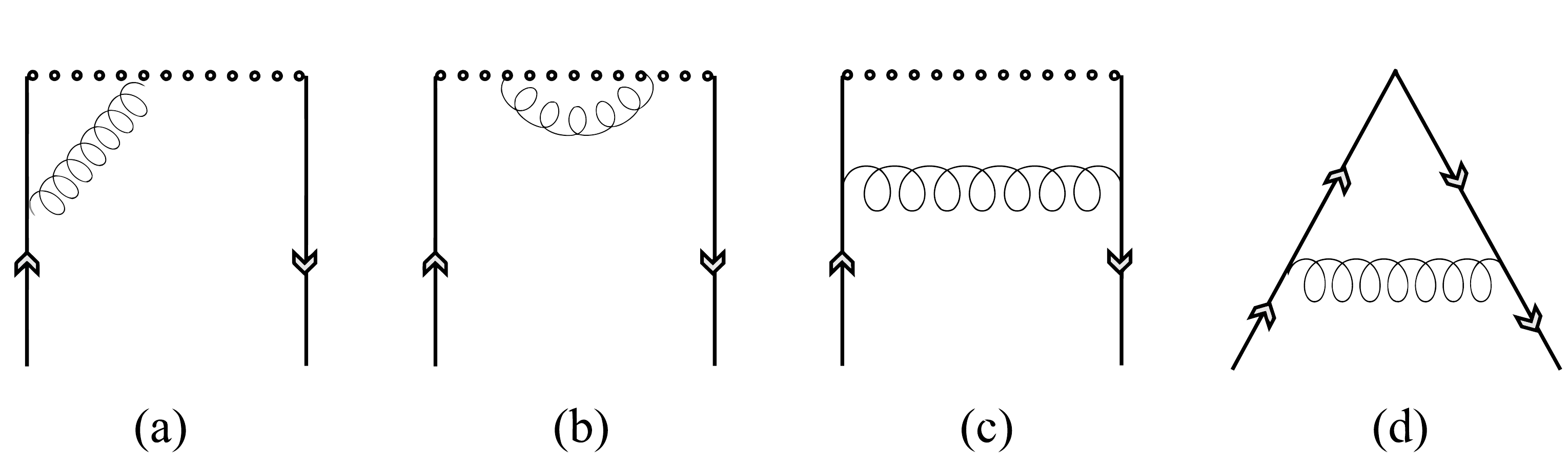}
  \caption{One-loop Feynman diagrams for the $B$-meson SFs. From left to right: (a) heavy-quark sail graph, (b) Wilson-line self-energy graph, (c) box graph, and (d) local vertex graph. The $b$ quark is denoted as solid line, while the Wilson line appears as dotted line.}
  \label{fig:feynman}
\end{figure*}

In the following, we will list the results of these graphs one by one. Starting from the quasi-SF $\tilde{S}^{\textrm{LaM.}}(x,\mu)$, its heavy-quark sail graph in Fig.\,\ref{fig:feynman}(a) gives:
\begin{align}\label{eq:quasi_sail}
  &\tilde{S}^{\textrm{LaM.}(1,a)}(x,\mu) = \begin{cases}
  \left[ -m_B\left( \frac{1}{m_b+\tilde{k}-xm_B} + \frac{xm_B}{(m_b+\tilde{k})(m_b+\tilde{k}-xm_B)} \ln\frac{(m_b+\tilde{k}-xm_B)^2}{(xm_B)^2} \right) \theta\Big(x-\frac{m_b}{m_B}-\frac{\tilde{k}}{m_B}\Big)\right]_\oplus\,,\\[10pt]
  \left[ -m_B \left( \frac{2 xm_B-m_b-\tilde{k}}{(m_b+\tilde{k}) (m_b+\tilde{k}-xm_B)} + \frac{xm_B}{(m_b+\tilde{k}) (m_b+\tilde{k}-xm_B)} \ln\frac{m_b^2(v \cdot k)^2(m_b+\tilde{k})^2}{4P^{z4}_B(xm_B)^2} \right) \theta\Big(\frac{m_b}{m_B}+\frac{\tilde{k}}{m_B}-x\Big)\theta(x) \right]_\oplus\,,\\[10pt]
  \left[ -m_B\left( -\frac{1}{m_b+\tilde{k}-xm_B} - \frac{xm_B}{(m_b+\tilde{k})(m_b+\tilde{k}-xm_B)} \ln\frac{(m_b+\tilde{k}-xm_B)^2}{(xm_B)^2} \right) \theta(-x)\right]_\oplus\,.
  \end{cases}
\end{align}

The Wilson-line self-energy graph in Fig.\,\ref{fig:feynman}(b) yields:
\begin{eqnarray}\label{eq:quasi_wilson}
  \tilde{S}^{\textrm{LaM.}(1,b)}(x,\mu) = \begin{cases}
  \left[ \frac{m_B}{xm_B-m_b-\tilde{k}} \, \theta\Big(x-\frac{m_b}{m_B}-\frac{\tilde{k}}{m_B}\Big) \right]_{\oplus} \,,\\[10pt]
  \left[ \frac{m_B}{m_b+\tilde{k}-xm_B} \, \theta\Big(\frac{m_b}{m_B}+\frac{\tilde{k}}{m_B}-x\Big) \right]_{\oplus} \,.
  \end{cases}
\end{eqnarray}

The result of box graph in Fig.\,\ref{fig:feynman}(c) reads:
\begin{align}\label{eq:quasi_box}
  &\tilde{S}^{\textrm{LaM.}(1,c)}(x,\mu) = \nn \\
  &\quad \begin{cases}
     \left[ m_B \left( \frac{1}{m_b+\tilde{k}} + \frac{m_b+\tilde{k}-xm_B}{(m_b+\tilde{k})^2} \ln\frac{-xm_B}{m_b+\tilde{k}-xm_B} \right) \theta(x-\frac{m_b}{m_B}-\frac{\tilde{k}}{m_B})\right]_\oplus \\
    + \int^{\infty}_{\frac{m_b}{m_B}+\frac{\tilde{k}}{m_B}} dt\, \left[  m_B \left(\frac{1}{m_b+\tilde{k}}+\frac{m_b+\tilde{k}-tm_B}{(m_b+\tilde{k})^2}\ln\frac{-tm_B}{m_b+\tilde{k}-tm_B}\right) \right]  \delta(x-\frac{m_b}{m_B}-\frac{\tilde{k}}{m_B})\,,\\[10pt]
     \left[ m_B \left( \frac{xm_Bm_b}{(m_b+\tilde{k})^2}\frac{1}{v\cdot k} + \frac{1}{(m_b+\tilde{k})^2}  \left(3xm_B-2m_b-2\tilde{k} + (m_b+\tilde{k}-xm_B) \ln\frac{2xm_B P^{z2}_B}{m_b(m_b+\tilde{k})(-v\cdot k)} \right) \right) \theta(\frac{m_b}{m_B}+\frac{\tilde{k}}{m_B}-x)\theta(x) \right]_\oplus\\
    +  \int_{0}^{\frac{m_b}{m_B}+\frac{\tilde{k}}{m_B}} dt \,\left[m_B\left( \frac{tm_B m_b}{(m_b+\tilde{k})^2}\frac{1}{ v\cdot k} + \frac{1}{(m_b+\tilde{k})^2}\left(3tm_B-2m_b-2\tilde{k}+(m_b+\tilde{k}-tm_B) \ln\frac{2tm_B P^{z2}_B}{m_b(m_b+\tilde{k})(-v\cdot k)}\right)\right) \right]\\
    \times \, \delta(x - \frac{m_b}{m_B} - \frac{\tilde{k}}{m_B})\,,\\[10pt]
    \left[ m_B \left( -\frac{1}{m_b+\tilde{k}} - \frac{m_b+\tilde{k}-xm_B}{(m_b+\tilde{k})^2} \ln\frac{-xm_B}{m_b+\tilde{k}-xm_B} \right) \theta(-x)\right]_\oplus \\
    +  \int_{-\infty}^0  dt\, \left[ m_B\left( -\frac{1}{m_b+\tilde{k}}-\frac{m_b+\tilde{k}-tm_B}{(m_b+\tilde{k})^2}\ln\frac{-tm_B}{m_b+\tilde{k}-tm_B}\right) \right]  \delta(x-\frac{m_b}{m_B}-\frac{\tilde{k}}{m_B})\,.
\end{cases}
\end{align}

The local vertex graph in Fig.\,\ref{fig:feynman}(d) gives:
\begin{align}\label{eq:quasi_local_vertex}
  & \tilde{S}^{\textrm{LaM.}(1,d)}(x,\mu) = \nn \\
  & \quad\begin{cases}
        \int^{\infty}_{\frac{m_b}{m_B}+\frac{\tilde{k}}{m_B}} dt \, \left[  m_B \left(\frac{1}{m_b+\tilde{k}}+\frac{m_b+\tilde{k}-tm_B}{(m_b+\tilde{k})^2}\ln\frac{-tm_B}{m_b+\tilde{k}-tm_B}\right) \right] \delta(x-\frac{m_b}{m_B}-\frac{\tilde{k}}{m_B})\,,\\[10pt]
        \int_{0}^{\frac{m_b}{m_B}+\frac{\tilde{k}}{m_B}} dt \,\left[m_B\left( \frac{tm_B m_b}{(m_b+\tilde{k})^2 }\frac{1}{v\cdot k} + \frac{1}{(m_b+\tilde{k})^2}\left(3tm_B-2\tilde{k}-2m_b+(m_b+\tilde{k}-tm_B) \ln\frac{2tm_B P^{z2}_B}{m_b(m_B+\tilde{k})(-v\cdot k)}\right)\right) \right] \\
        \times \, \delta(x-\frac{m_b}{m_B}-\frac{\tilde{k}}{m_B}) \,,\\[10pt]
        \int_{-\infty}^0  dt \,\left[ m_B\left( -\frac{1}{m_b+\tilde{k}}-\frac{m_b+\tilde{k}-tm_B}{(m_b+\tilde{k})^2}\ln\frac{-tm_B}{m_b+\tilde{k}-tm_B}\right) \right] \delta(x-\frac{m_b}{m_B}-\frac{\tilde{k}}{m_B})\,.
        \end{cases}
\end{align}

Eqs.\,(\ref{eq:quasi_sail})-(\ref{eq:quasi_local_vertex}) constitute the one-loop results for the quasi-SF. One can observe that the IR divergences $\ln(-v\cdot k)$ appear in the individual diagram results. As we shall see below, these will cancel in the matching procedure, rendering the matching coefficient free of IR dependence. Furthermore, the explicit dependence on the heavy quark mass also cancels in the matching, ensuring that the matching coefficient is also independent of $m_b$, as discussed in the previous section.

\subsection{One-loop results for the shape function $S^{\textrm{QCD}}$ in QCD}
\label{sec:oneloop_QCD}

We now present the one-loop results for the QCD SF $S^{\textrm{QCD}}(y,\mu)$. The result of the heavy-quark sail graph in Fig.\,\ref{fig:feynman}(a) is:
\begin{eqnarray}\label{eq:QCD_sail}
  && S^{\textrm{QCD}(1,a)}(y,\mu) =  \left[\frac{2ym_B^2}{(m_b+\tilde{k}-ym_B)(m_b+\tilde{k})} \left( \frac{1}{\epsilon} + \ln\frac{(m_b+\tilde{k})\mu^2}{2m_b (m_b+\tilde{k}-ym_B) (-v\cdot k)} \right) \theta\Big(\frac{m_b}{m_B}+\frac{\tilde{k}}{m_B}-y\Big)\theta(y) \right]_\oplus \,.
\end{eqnarray}

The contribution of the Wilson-line self-energy graph in Fig.\,\ref{fig:feynman}(b) vanishes for the QCD SF, since the amplitude is proportional to $n_+^2=0$.

The result of box graph in Fig.\,\ref{fig:feynman}(c) is:
\begin{eqnarray}\label{eq:QCD_box}
  &&S^{\textrm{QCD}(1,c)}(y,\mu) = \nonumber\\
  &&\frac{m_B}{m_b+\tilde{k}} \Bigg[ \left(\frac{m_b+\tilde{k}-ym_B}{m_b+\tilde{k}} \frac{1}{\epsilon} + \frac{ym_Bm_b}{m_b+\tilde{k}}\frac{1}{v\cdot k} + \frac{m_b+\tilde{k}-ym_B}{m_b+\tilde{k}} \left( -2 + \ln\frac{(m_b+\tilde{k})\mu^2}{2m_b(m_b+\tilde{k}-ym_B)(-v\cdot k)} \right) \right) \nn\\
  &&\times \theta\Big(\frac{m_b}{m_B}+\frac{\tilde{k}}{m_B}-y\Big)\theta(y)\Bigg]_\oplus \nn\\
  && + \frac{m_B}{m_b+\tilde{k}} \int_{0}^{\frac{m_b}{m_B}+\frac{\tilde{k}}{m_B}} dt\, \left[\frac{m_b+\tilde{k}-tm_B}{m_b+\tilde{k}} \frac{1}{\epsilon} + \frac{tm_Bm_b}{m_b+\tilde{k}}\frac{1}{v\cdot k} + \frac{m_b+\tilde{k}-tm_B}{m_b+\tilde{k}} \left(-2 + \ln\frac{(m_b+\tilde{k})\mu^2}{2m_b(m_b+\tilde{k}-tm_B)(-v\cdot k)} \right) \right] \nn\\
  && \times \delta\Big(y-\frac{m_b}{m_B} - \frac{\tilde{k}}{m_B}\Big) \,.
\end{eqnarray}

The result of local vertex graph in Fig.\,\ref{fig:feynman}(d) is:
\begin{align}\label{eq:QCD_local_vertex}
  &S^{\textrm{QCD}(1,d)}(y,\mu) = \nonumber \\
  &\frac{m_B}{m_b+\tilde{k}} \int_{0}^{\frac{m_b}{m_B}+\frac{\tilde{k}}{m_B}} dt\, \left[\frac{m_b+\tilde{k}-tm_B}{m_b+\tilde{k}} \frac{1}{\epsilon} + \frac{tm_Bm_b}{m_b+\tilde{k}}\frac{1}{v\cdot k} + \frac{m_b+\tilde{k}-tm_B}{m_b+\tilde{k}} \left( -2 + \ln\frac{(m_b+\tilde{k})\mu^2}{2m_b(m_b+\tilde{k}-tm_B)(-v\cdot k)} \right) \right] \nn\\
  & \times \delta\Big(y-\frac{m_b}{m_B} - \frac{\tilde{k}}{m_B}\Big) \,.
\end{align}

Eqs.\,(\ref{eq:QCD_sail})-(\ref{eq:QCD_local_vertex}) constitute the one-loop results for the QCD SF. Similar to the quasi-shape function case, the IR divergences $\ln(-v\cdot k)$ appear in the individual diagram results. These will cancel when forming the matching.

\subsection{One-loop results for the matching coefficient}
\label{sec:oneloop_matching}

Based on the above results of quasi-SF and QCD SF, as well as the one-loop matching formula, which is derived from the factorization formula in Eq.\,(\ref{eq:fac1})
\begin{eqnarray}
  Z_\Gamma^{(1)}(x,y,\mu) = \left.\left[ \tilde{S}^{\textrm{LaM.}(1)}(x,\mu) -  S^{\textrm{QCD}(1)}(y,\mu) \big|_{y \to x}  \right]\right|_{\frac{m_b}{m_B}+\frac{\tilde{k}}{m_B} \to y} \,,
\end{eqnarray}
the matching coefficient at one-loop can be obtained:
\begin{equation}\label{eq:matching_final}
    \begin{aligned}
         Z_\Gamma^{(1)}(x,y,\mu)
         =\begin{cases}
        \left[ \frac{1}{y^2(y-x)} \left( y(x+y)+(x^2+y^2)\ln \frac{x-y}{x}  \right) \right]_\oplus \theta(-x) \,,  \\[10pt]
        \left[ \frac{1}{y^2(y-x)}\left( 2y^2-xy-x^2-(x^2+y^2)\ln \frac{y^2\mu^2}{4P_B^{z2}x(y-x)}     \right)  \right]_\oplus \theta(y-x)\theta(x)\,,  \\[10pt]
        \left[ \frac{1}{y^2(y-x)} \left( -y(x+y)-(x^2+y^2)\ln \frac{x-y}{x}  \right) \right]_\oplus \theta(x-y) \,.
        \end{cases}
    \end{aligned}
\end{equation}
Remarkably, the matching coefficient retains no dependence on the heavy quark or heavy meson mass, which aligns precisely with the idea of scale separation in the HQLaMET framework. The above expression can be written more compactly by introducing $\xi=x/y$, which yields:
\begin{eqnarray}\label{eq:matching_finalxi}
  Z_\Gamma^{(1)}(\xi,y,\mu) = \begin{cases}
  \dfrac{1}{y}\left[\dfrac{1}{\xi-1}\left( -1-\xi+(1+\xi^2)\ln\dfrac{\xi}{\xi-1}  \right)\right]_\oplus \theta(-\xi) \,,\\[10pt]
  \dfrac{1}{y}\left[\dfrac{1}{\xi-1}\left( -2+\xi+\xi^2+(1+\xi^2)\ln\dfrac{\mu^2}{4P_B^{z2}(1-\xi)\xi}  \right)\right]_\oplus \theta(1-\xi)\theta(\xi) \,, \\[10pt]
  \dfrac{1}{y}\left[\dfrac{1}{\xi-1}\left( 1+\xi -(1+\xi^2)\ln\dfrac{\xi}{\xi-1}  \right) \right]_\oplus \theta(\xi-1) \,.
  \end{cases}
\end{eqnarray}

\end{widetext}
The one-loop matching coefficient $Z_\Gamma^{(1)}(\xi,y,\mu)$ in Eq.\,(\ref{eq:matching_finalxi}) presents one of the main results of this work. Several features are noteworthy. First, the result is free of any IR divergences, confirming the validity of the factorization formula. Second, the matching coefficient depends solely on the large momentum $P_B^z$ and is independent of the heavy quark mass $m_b$. This is consistent with the design of the two-step factorization scheme: the first step integrates out only the hard scale $P_B^z$, while the heavy quark mass $m_b$ is carried entirely from the quasi-SF into the QCD SF and would be separated out subsequently in the second step.

It should be emphasized that, the matching coefficient obtained in the $v\cdot k\neq 0$ scheme is identical to the result obtained in the $m_g\neq 0$ scheme (\app{appendixA}). This serves as a nontrivial cross-check of our calculation.

\section{Matching relation between $S^{\textrm{QCD}}$ and $S^{\textrm{HQET}}$ at next-to-leading order}
\label{sec:secondstep}

The QCD SF incorporates dependence on the heavy quark mass and encapsulates hadronic physics at two distinct scales, $m_b$ and $\Lambda_{\textrm{QCD}}$. To separate these scales, the second step factorization connects $S^{\textrm{QCD}}(y,\mu)$ to $S^{\textrm{HQET}}(\omega,\mu)$ through the matching formula given in Eq.\,(\ref{eq:fac2}). Intuitively, one expects the heavy quark to carry most of the lightcone momentum in the $B$-meson, localizing the SF near $y\sim 1$. In this peak region, where $1-y \sim \Lambda_{\textrm{QCD}}/m_b$, the factorization reduces from a convolution to a simple multiplication~\cite{Wang:2025uap}. In contrast, the tail region where $1-y \sim \mathcal{O}(1)$ is fully perturbative.

As derived in Ref.\,\cite{Wang:2025uap}, the one-loop matching coefficient in the peak region takes a remarkably simple form:
\begin{eqnarray}\label{eq:peak_matching}
  && \left. S^{\textrm{QCD}}(y,\mu) \right|_{\textrm{peak}} \nn\\
  &&= \left[ 1 + \frac{\alpha_s C_F}{2\pi}\left( \frac{1}{2}\ln^2\frac{\mu^2}{m_b^2} -\frac{3}{2}\ln\frac{\mu^2}{m_b^2} +\frac{\pi^2}{12} -2 \right) \right] \nn\\
  && \quad \times \left. S^{\textrm{HQET}}(\omega,\mu) \right|_{\omega \to ym_B -m_b}  \,,
\end{eqnarray}
with
\begin{eqnarray}
  Z_{\textrm{peak}}^{(1)}(y,\omega,\mu) &=& \left( \frac{1}{2}\ln^2\frac{\mu^2}{m_b^2} -\frac{3}{2}\ln\frac{\mu^2}{m_b^2} +\frac{\pi^2}{12} -2 \right) \nn\\
  &&\times \, \delta\Big(y-\frac{m_b}{m_B}-\frac{\omega}{m_B} \Big) \,.
\end{eqnarray}

Through the HQLaMET framework, the effects from three disparate scales---$P_B^z$, $m_b$, and $\Lambda_{\textrm{QCD}}$---are fully disentangled: $P_B^z$ enters the first step matching coefficient $Z_\Gamma(x,y,\mu)$, $m_b$ enters the second step matching coefficient $Z_{\textrm{peak}}(y,\omega,\mu)$, and $\Lambda_{\textrm{QCD}}$ is encoded in the nonperturbative HQET SF $S^{\textrm{HQET}}(\omega,\mu)$.

\section{Mass dependence of $S^{\textrm{QCD}}$}
\label{sec:massevolution}

The QCD SF depends on two scales: the factorization scale $\mu$ and the heavy quark mass $m_Q$. The scale dependence is governed by the standard DGLAP evolution equation; here we focus on the mass dependence. Understanding the heavy quark mass dependence is of particular interest for lattice calculations, where simulations at the physical $b$ quark mass may be computationally prohibitive and one must instead evolve results obtained at lower masses (e.g., charm quark mass) to the physical point. This constitutes the central objective of the present section.

In the following derivation, we will explicitly write down the dependence of the QCD SF on the heavy quark mass. The definition of $S^{\textrm{QCD}}(y,m_Q,\mu)$ has been given in Eq.\,(\ref{eq:def_QCD_SF}), where the mass dependence enters through the heavy quark field. In this work the heavy-quark mass entering the peak-region matching and the variable mapping is understood as the pole mass. The resulting mass-dependence equation is therefore a fixed-order, leading-power relation in the pole-mass scheme. For precision phenomenology, the pole-mass renormalon should be removed by converting to a short-distance mass scheme, which induces a corresponding shift of the HQET shape-function argument.

In the peak region, the QCD SF is related to the HQET SF through:
\begin{eqnarray}\label{eq:mass_matching}
  S^{\textrm{QCD}}(y,m_Q,\mu)\big|_{\rm peak} = J_S(m_Q,\mu)\, S^{\textrm{HQET}}(\omega,\mu) \,,
\end{eqnarray}
where the matching coefficient is
\begin{eqnarray}\label{eq:JS}
  J_S(m_Q,\mu) &=& 1+ \frac{\alpha_s C_F}{2\pi} \bigg( \frac{1}{2} \ln^2\frac{\mu^2}{m_Q^2} - \frac{3}{2} \ln\frac{\mu^2}{m_Q^2}  + \frac{\pi^2}{12} -2 \bigg) \nn\\
  && +\, \mathcal{O}(\alpha_s^2) \,,
\end{eqnarray}
and the variable mapping is $\omega \simeq \bar\Lambda-(1-y)m_Q$ at leading power in $\Lambda_{\textrm{QCD}}/m_Q$. Here $m_Q$ denotes the heavy quark pole mass and $\bar\Lambda = m_H - m_Q$ with $m_H$ being the heavy meson mass.

Since the HQET SF depends on $\omega$ but not on $m_Q$ explicitly, one can construct a differential operator that preserves $\omega$. Noting that
\begin{eqnarray}
  m_Q \frac{\partial \omega}{\partial m_Q} + (1-y)\frac{\partial \omega}{\partial y} = 0 \,,
\end{eqnarray}
we define
\begin{eqnarray}\label{eq:operator_B}
  \hat{\mathcal{O}} = m_Q\frac{\partial}{\partial m_Q} + (1-y)\frac{\partial}{\partial y} \,.\nn
\end{eqnarray}
Applying $\hat{\mathcal{O}}$ to the matching formula Eq.\,(\ref{eq:mass_matching}), and using $\hat{\mathcal{O}}\,\omega = 0$ together with the fact that $J_S(m_Q,\mu)$ depends only on $m_Q$ and $\mu$, one obtains the mass dependence equation:
\begin{eqnarray}\label{eq:mass_PDE}
  && \bigg[ m_Q\frac{\partial}{\partial m_Q} + (1-y)\frac{\partial}{\partial y} - \gamma_S(m_Q,\mu) \bigg] \nn\\
  && \quad \times\, S^{\textrm{QCD}}(y,m_Q;\mu) = 0 \,,
\end{eqnarray}
where
\begin{eqnarray}\label{eq:Gamma_def}
  \gamma_S(m_Q,\mu) \equiv \frac{\partial \ln J_S(m_Q,\mu)}{\partial \ln m_Q} \,.\nn
\end{eqnarray}

At fixed order in $\alpha_s$, we obtain
\begin{eqnarray}\label{eq:Gamma_fixed}
  \gamma_S(m_Q,\mu) = \frac{\alpha_s C_F}{2\pi}\left( 3 - 4\ln\frac{\mu}{m_Q} \right) \,.
\end{eqnarray}
After RGE improvement, which resums potentially large logarithms between $\mu$ and $m_Q$, the result becomes
\begin{eqnarray}\label{eq:Gamma_RGE}
  \gamma_S(m_Q,\mu) \simeq \frac{3 \alpha_s(m_Q) C_F}{2\pi} + \frac{4C_F}{\beta_0} \ln\frac{\alpha_s(\mu)}{\alpha_s(m_Q)} \,,
\end{eqnarray}
where $\beta_0 = 11 - 2n_f/3$.

The mass dependence in Eq.\,(\ref{eq:mass_PDE}) is a first-order linear partial differential equation (PDE) that can be solved by the method of characteristics. The characteristic lines satisfy $m_Q(1-y) = \textrm{const}$, giving the initial-point relation:
\begin{eqnarray}\label{eq:characteristic}
  y_0 = 1 - \frac{m_Q}{m_{Q_0}}(1-y) \,.
\end{eqnarray}
Along the characteristics, the solution is:
\begin{eqnarray}
S^{\mathrm{QCD}}(y,m_Q,\mu)& = &\exp\!\left[\int_{m_{Q_0}}^{m_Q}\frac{dm'}{m'}\,\gamma_S(m',\mu)\right]\nn\\
&&\hspace{-1.5 em}\times\frac{m_Q}{m_{Q_0}}\,S^{\mathrm{QCD}}\!\left(y_0,m_{Q_0},\mu\right) .
\label{eq:mass_solution}
\end{eqnarray}

If we expand to fixed order with $\alpha_s$ evaluated at $m_{Q_0}$, defining $R\equiv \ln(m_Q/m_{Q_0})$ and $L_0\equiv \ln(\mu/m_{Q_0})$, the solution takes the form:

\begin{eqnarray}
S^{\mathrm{QCD}}(y,m_Q,\mu)
&=&
\left[1+\frac{\alpha_s(m_{Q_0})C_F}{2\pi}\left(2R^2-4L_0R+3R\right)\right]\nn\\
&&\hspace{-3 em}\times\,\frac{m_Q}{m_{Q_0}}\,S^{\mathrm{QCD}}\left(1-\frac{m_Q}{m_{Q_0}}(1-y),m_{Q_0},\mu\right)\, .
\label{eq:mass_solution_expanded}
\end{eqnarray}
This result allows one to evolve the QCD SF from an initial mass $m_{Q_0}$ (where lattice data or model input is available) to a different mass $m_Q$, with the argument of the initial SF being rescaled according to the characteristic solution.

We remark that the mass dependence derived here applies to the peak region at leading power in $\Lambda_{\textrm{QCD}}/m_Q$. In the tail region where $1-y\sim\mathcal{O}(1)$, the SF can be computed perturbatively and its mass dependence is explicit. A unified treatment connecting the peak and tail regions requires incorporating subleading power corrections and is left for future study.

\section{Perspectives for lattice calculations}
\label{sec:numerical}

The SF is a nonperturbative object that has not been predicted from first-principles up to now. However, valuable phenomenological models exist \cite{Bosch:2004th,Balzereit:1998yf,Ligeti:2008ac} and can serve as starting points for illustrating the full HQLaMET framework. In this section, we provide a numerical demonstration by starting from an HQET SF model and working backwards through the complete chain of this framework.

Here, we adopt the widely used parametric model for the HQET SF at the soft scale $\mu_i = 1.5\,\textrm{GeV}$~\cite{Bosch:2004th}:
\begin{eqnarray}\label{eq:model}
  && \hat{S}^{\textrm{HQET}}(\hat{\omega}, \mu)=\frac{N}{A}\left(\frac{\hat{\omega}}{A}\right)^{b-1} \exp \left(-b \frac{\hat{\omega}}{A}\right) \nn\\
  && -\frac{\alpha_s C_F}{\pi} \frac{\theta(\hat{\omega}-A-\mu / \sqrt{e})}{\hat{\omega}-A}  \left( 2 \ln \frac{\hat{\omega}-A}{\mu} +1 \right) ,
\end{eqnarray}
with the normalization factor
\begin{eqnarray}
  N=\left[1-\frac{\alpha_s C_F}{\pi}\left(\frac{\pi^2}{24}-\frac{1}{4}\right)\right] \frac{b^b}{\Gamma(b)} \,.
\end{eqnarray}
The model parameters at the intermediate scale $\mu_i=1.5\,\textrm{GeV}$ are $A=0.685\,\textrm{GeV}$ and $b=2.93$. The variable $\hat{\omega}=\bar\Lambda-\omega \geq 0$  corresponds to the momentum fraction via the relation $\hat{\omega}=(1-y)m_B$. We take $m_B=5.28\,\textrm{GeV}$, $m_b = 4.78\,\textrm{GeV}$, and $m_c = 1.67\,\textrm{GeV}$. The strong coupling constant is evaluated using the three-loop running (from \textsc{RunDec} \cite{Herren:2017osy}).

\begin{figure}[!htbp]
  \centering
  \includegraphics[width=0.95\columnwidth]{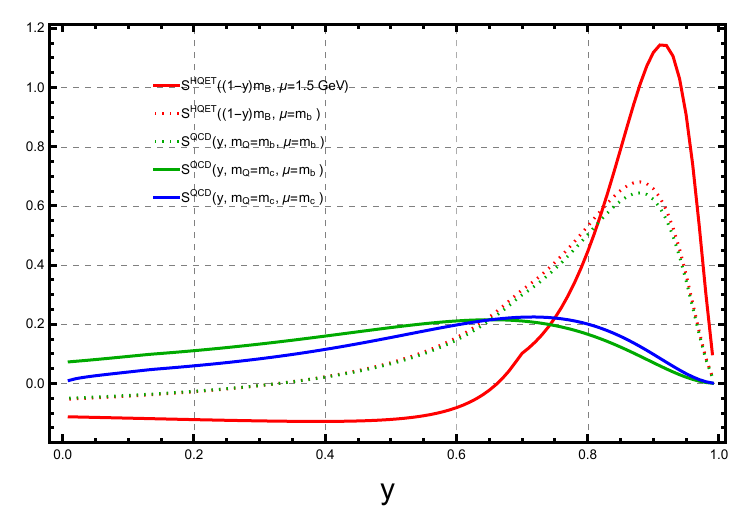}
  \caption{Illustration of the second step of the HQLaMET framework. The red solid curve is model of HQET SF given at $\mu_0=1.5\,\textrm{GeV}$; The red dotted curve is HQET SF evolved to $\mu=m_b$; The green dotted curve is QCD SF matched at $m_Q=m_b$, $\mu=m_b$; The green solid curve is QCD SF evolved to $m_Q=m_c$, $\mu=m_b$; The blue solid curve is the QCD SF evolved to $\mu=m_c$ with $m_Q=m_c$.}
  \label{fig:step2}
\end{figure}

We start with the HQET SF model at the soft scale $\mu_i=1.5\,\textrm{GeV}$ and evolve it within HQET to the matching scale $\mu = m_B$. The solution to the evolution equation for the HQET SF is \cite{Bosch:2004th}:
\begin{eqnarray}\label{eq:LN_evolution}
  S^{\textrm{HQET}}(\hat{\omega},\mu_i) &=& e^{V_S(\mu_i,\mu_0)}\frac{1}{\Gamma(\eta)}  \int_0^{\hat{\omega}} d\hat{\omega}' \, \frac{S^{\textrm{HQET}}(\hat{\omega}' ,\mu_0)}{\mu_0^\eta(\hat{\omega}-\hat{\omega}')^{1-\eta}} \,,\nn
\end{eqnarray}
where $V_S$ and $\eta$ are defined through the anomalous dimension and the cusp anomalous dimension; for their explicit expressions we refer the reader to Ref.\,\cite{Bosch:2004th}. The evolution result is illustrated in Fig.\,\ref{fig:step2}, where the red solid curve shows the HQET SF at the initial scale $\mu_0=1.5\,\textrm{GeV}$ and the red dotted curve shows the evolved result at $\mu=m_b$.

Utilizing the second step matching result from Eq.\,(\ref{eq:peak_matching}), the HQET SF at $\mu=m_b$ (red dotted curve in Fig.\,\ref{fig:step2}) is matched onto the QCD SF at $m_Q=m_b$, $\mu=m_b$ (green dotted curve in Fig.\,\ref{fig:step2}). Subsequently, the mass dependence solution in Eq.\,(\ref{eq:mass_solution_expanded}) is applied to evolve from $m_Q=m_b$ down to $m_Q=m_c$, yielding the QCD SF with $m_Q=m_c$, $\mu=m_b$ (green solid curve in Fig.\,\ref{fig:step2}).

\begin{figure}[!htbp]
  \centering
  \includegraphics[width=0.95\columnwidth]{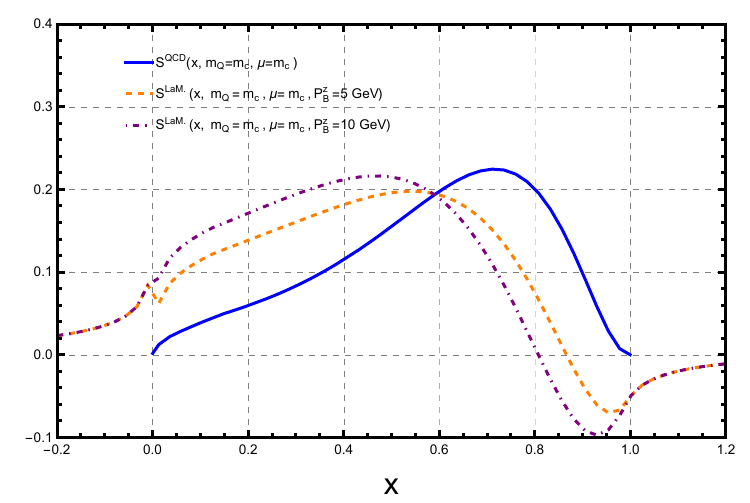}
  \caption{The quasi-SFs obtained from the first step matching at two values of $P_B^z$. The quasi-SFs (dashed curves) exhibit a clear dependence on $P_B^z$. Notably, they are nonzero in the unphysical regions ($x<0$ and $x>1$).}
  \label{fig:step1}
\end{figure}

The scale evolution of the QCD SF is governed by the standard DGLAP equation. Employing this equation, we evolve the QCD SF from $\mu=m_b$ (green solid curve in Fig.\,\ref{fig:step2}) to $\mu=m_c$ (blue solid curve in Fig.\,\ref{fig:step2}), thereby completing the final preparation for matching to the quasi-SF. Finally, applying the first step matching result from Eq.\,(\ref{eq:matching_finalxi}), we obtain the quasi-SFs from the QCD SF at $m_Q=m_c$, $\mu=m_c$. The results for two representative values of $P_B^z$ are displayed in Fig.\,\ref{fig:step1}. As shown in Fig.\,\ref{fig:step1}, the quasi-SFs exhibit a clear dependence on the large momentum $P_B^z$, which is a characteristic feature of the HQLaMET framework. Furthermore, they develop nonzero support in the unphysical regions $x<0$ and $x>1$, analogous to the behavior observed in quasi-distribution amplitudes. One may anticipate that future lattice simulations of the quasi-SF will display similar features.

Figs.\,\ref{fig:step2} and \ref{fig:step1} together constitute a model-dependent numerical illustration of the HQLaMET framework for the $B$-meson SF. While the starting point relies on a phenomenological model, the entire chain of operations---the scale evolution in HQET, second step matching, mass dependence, DGLAP evolution, and first step matching---is built on rigorous theoretical analyses. This numerical illustration will serve as a valuable guide for future lattice simulations, which will eventually replace the model with first-principles nonperturbative results for the HQET SF, thereby facilitating future phenomenological studies of inclusive $B$-meson decays.

\section{Summary}
\label{sec:summary}

The shape function defined in HQET describes the Fermi motion of the heavy $b$ quark inside the $B$ meson, encoding the distribution of its residual lightcone momentum arising from strong interactions. It serves as a fundamental nonperturbative input in precision studies of inclusive $B$ decays, playing a role analogous to that of parton distribution functions in nucleon physics. However, the long-standing model dependence of this essential quantity has remained a major limitation in phenomenological studies.

In this work, we have extended the HQLaMET framework to facilitate future lattice computations of the $B$-meson shape function. This framework allows one to start from a quasi-shape function computable on the lattice and, through a systematic sequence of operations, obtain the desired HQET shape function. The key results of this paper are: (i) the one-loop matching coefficient $Z_\Gamma(\xi,y,\mu)$ between the quasi-shape function and the QCD shape function, given in Eq.\,(\ref{eq:matching_finalxi}); (ii) the mass dependence equation and its analytic solution, given in Eqs.\,(\ref{eq:mass_PDE}) and (\ref{eq:mass_solution}); and (iii) a numerical illustration of the complete HQLaMET workflow. These results lay the theoretical foundation for first-principles lattice QCD determinations of the $B$-meson shape function in the near future.

\begin{widetext}
\appendix
\section{Results with gluon mass as IR regulator}\label{sec:appendixA}

In this appendix, we present the one-loop results for the quasi-SF and QCD SF obtained with an alternative infrared regulator, namely a nonzero gluon mass $m_g\neq 0$. Here, the heavy quark mass is set to zero. The calculation in this appendix provides a nontrivial cross-check of the results presented in the main text.

\subsection{One-loop results for the quasi-shape function $\tilde{S}^{\textrm{LaM.}}$ in LaMET with nonzero gluon mass}

Here, we present directly the one-loop results for the individual graphs in Fig.\,\ref{fig:feynman} for the quasi-SF. The heavy-quark sail graph in Fig.\,\ref{fig:feynman}(a) is:
\begin{eqnarray}\label{eq:quasi_sail_mg}
  \tilde{S}^{\textrm{LaM.}(1,a)}(x,\mu)\big|_{m_g} = \begin{cases}
  \left[-\dfrac{1-2x\ln\frac{x}{x-1}}{1-x} \, \theta(x-1)\right]_\oplus\,,\\[10pt]
  \left[-\dfrac{2x-1-2x\ln\frac{4P_B^{z2}(1-x)}{m_g^2}}{1-x} \, \theta(x)\theta(1-x)\right]_\oplus \,,\\[10pt]
  \left[\dfrac{1-2x\ln\frac{x}{x-1}}{1-x} \, \theta(-x)\right]_\oplus\,.
  \end{cases}
\end{eqnarray}

The Wilson-line self-energy graph in Fig.\,\ref{fig:feynman}(b) yields:
\begin{eqnarray}\label{eq:quasi_wilson_mg}
  \tilde{S}^{\textrm{LaM.}(1,b)}(x,\mu)\big|_{m_g} = \begin{cases}
  \left[ \dfrac{1}{x-1}\theta(x-1) \right]_{\oplus} \,,\\[10pt]
  \left[ \dfrac{1}{1-x}\theta(1-x) \right]_{\oplus} \,.
  \end{cases}
\end{eqnarray}

The box graph in Fig.\,\ref{fig:feynman}(c):
\begin{eqnarray}\label{eq:quasi_box_mg}
  \tilde{S}^{\textrm{LaM.}(1,c)}(x,\mu)\big|_{m_g}= \begin{cases}
        \left[  \left( 1 - (1-x) \ln\frac{x-1}{x} \right) \theta(x-1)\right]_\oplus + \int_0^1 dt \,\left[1 - (1-t)\ln\frac{t-1}{t} \right]  \delta(x-1) \,,\\[10pt]
        \left[  \left(3x-2 - (1-x) \ln\frac{m_g^2}{4P_B^{z2}(1-x)} \right)  \theta(x)\theta(1-x) \right]_\oplus
        +\int_0^{1} dt\, \left[3t-2 - (1-t)\ln\frac{m_g^2}{4P_B^{z2} (1-t)} \right] \\ \times \,  \delta(x-1) \,,\\[10pt]
        \left[  \left( -1 + (1-x) \ln\frac{x-1}{x} \right) \theta(-x)\right]_\oplus +\int_0^1 dt\, \left[-1 + (1-t)\ln\frac{t-1}{t} \right]  \delta(x-1) \,.
        \end{cases}
\end{eqnarray}

The local vertex graph in Fig.\,\ref{fig:feynman}(d):
\begin{eqnarray}\label{eq:quasi_local_vertex_mg}
   \tilde{S}^{\textrm{LaM.}(1,d)}(x,\mu)\big|_{m_g} = \begin{cases}
            \int_0^1 dt \,\left[1 - (1-t)\ln\frac{t-1}{t} \right]  \delta(x-1) \,,\\[10pt]
            \int_0^{1} dt \,\left[3t-2 - (1-t)\ln\frac{m_g^2}{4P_B^{z2} (1-t)} \right]  \delta(x-1) \,,\\[10pt]
            \int_0^1 dt \,\left[-1 + (1-t)\ln\frac{t-1}{t} \right]  \delta(x-1) \,.\\
        \end{cases}
\end{eqnarray}
Eqs.\,(\ref{eq:quasi_sail_mg})-(\ref{eq:quasi_local_vertex_mg}) constitute the one-loop results for the quasi-SF.

\subsection{One-loop results for the shape function $S^{\textrm{QCD}}$ in QCD with nonzero gluon mass}

We present the one-loop results for the individual graphs in Fig.\,\ref{fig:feynman} for the QCD SF. The heavy-quark sail graph in Fig.\,\ref{fig:feynman}(a) is:
\begin{eqnarray}\label{eq:QCD_sail_mg}
  S^{\textrm{QCD}(1,a)}(y,\mu)\big|_{m_g} = \left[\frac{2y}{1-y}  \left(  \frac{1}{\epsilon} + \ln\frac{\mu^2 }{m_g^2 \, y} \right) \theta(y)\theta(1-y)\right]_\oplus \,.
\end{eqnarray}

The Wilson-line self-energy contribution vanishes, since the amplitude is proportional to $n_+^2=0$.

The box graph in Fig.\,\ref{fig:feynman}(c):
\begin{eqnarray}\label{eq:QCD_box_mg}
  S^{\textrm{QCD}(1,c)}(y,\mu)\big|_{m_g} = \left[ (1-y) \left( \frac{1}{\epsilon} + \ln\frac{\mu^2}{y m_g^2}-2  \right) \theta(y)\theta(1-y) \right]_\oplus  + \delta(y-1) \int_0^1 dt \,(1-t) \left( \frac{1}{\epsilon} + \ln\frac{\mu^2}{t m_g^2}-2  \right) \,.
\end{eqnarray}

The local vertex graph in Fig.\,\ref{fig:feynman}(d):
\begin{eqnarray}\label{eq:QCD_local_vertex_mg}
  S^{\textrm{QCD}(1,d)}(y,\mu)\big|_{m_g} = \delta(y-1) \int_0^{1} dt\, (1-t) \left( \frac{1}{\epsilon} + \ln\frac{\mu^2}{tm_g^2}-2  \right) \,.
\end{eqnarray}
Eqs.\,(\ref{eq:QCD_sail_mg})-(\ref{eq:QCD_local_vertex_mg}) constitute the one-loop results for the QCD SF.

\subsection{One-loop results for the matching coefficient}
With the above results at hand, and following the same procedure as in the main text, we now present the one-loop matching coefficient in this $m_g\neq 0$ scheme (with $y\to 1$):
\begin{eqnarray}\label{eq:matching_mg}
  Z_\Gamma^{(1)}(x,y\to 1,\mu)\big|_{m_g} = \begin{cases}
  \left[\dfrac{1}{x-1}\left( -1-x+(1+x^2)\ln\dfrac{x}{x-1}  \right)\right]_\oplus \theta(-x) \,,\\[10pt]
  \left[\dfrac{1}{x-1}\left( -2+x+x^2+(1+x^2)\ln\dfrac{\mu^2}{4P_B^{z2}(1-x)x}  \right)\right]_\oplus \theta(x)\theta(1-x) \,, \\[10pt]
  \left[\dfrac{1}{x-1}\left( 1+x-(1+x^2)\ln\dfrac{x}{x-1}  \right) \right]_\oplus \theta(x-1) \,.
  \end{cases}
\end{eqnarray}
This result no longer depends on the gluon mass $m_g$, confirming the regulator independence of the matching coefficient. One can verify that setting $y\to 1$ in Eq.\,(\ref{eq:matching_finalxi}) yields precisely the same expression as Eq.\,(\ref{eq:matching_mg}), thereby providing a nontrivial validation of the calculation. The agreement between two independent IR regularization schemes demonstrates the robustness of the HQLaMET factorization framework.
\end{widetext}

\section*{Acknowledgements}
The authors would like to thank Profs. Wei Wang and Fu-Sheng Yu for inspiring discussions. This work was supported in part by the National Natural Science Foundation of China under Grant Nos. 12475098, 12565014, 12105247. The work of J.Z was also supported by the Talent Research Startup Foundation of Hainan Normal University: HSZK-KYQD-202523, Opening Foundation of Shanghai Key Laboratory of Particle Physics and Cosmology under Grant No. 22DZ2229013-5, and supported by Hainan Provincial Natural Science Foundation of China under Grant No. 126MS0134. J.X was supported in part by the Key Laboratory for Particle Astrophysics and Cosmology, Ministry of Education (MoE). A.S.X was supported by the Fundamental Research Funds for the Central Universities under No.~lzujbky-2023-stlt01.

\end{document}